\documentclass[prd,twocolumn,floatfix,amsmath,nofootinbib,amssymb,floatfix]{revtex4}
\usepackage{graphicx,color,dcolumn,booktabs,bm}
\usepackage{longtable,lscape}
\usepackage{txfonts}
\usepackage{overpic}
\usepackage{amssymb}
\usepackage{indentfirst}
\usepackage{feynmf}   
\usepackage{slashed}  
\usepackage{cases}
\usepackage{color}
\usepackage{multirow}
\usepackage{epstopdf}
\usepackage{enumerate}
\usepackage{graphicx,color,dcolumn,booktabs,bm}
\usepackage[colorlinks, citecolor=blue,anchorcolor=red,menucolor=red, linkcolor=red,filecolor=red,urlcolor=blue,frenchlinks=red]{hyperref}

\newcommand{\VEV}[1]{\left\langle #1 \right\rangle}

\begin{document}

\title{Early kinetic decoupling effect on the forbidden dark matter annihilations into standard model particles}
\author{Yu Liu}\email{liuyu2020@s.ytu.edu.cn}

\author{Xuewen Liu}\email{xuewenliu@ytu.edu.cn}

\author{Bin Zhu}\email{zhubin@mail.nankai.edu.cn}
\affiliation{Department of Physics, Yantai University, Yantai 264005, China}

\date{\today}

\begin{abstract}

The early kinetic decoupling (eKD) effect is an inevitable ingredient in 
calculating the relic density of dark matter (DM) for various well-motivated scenarios. It appears naturally in forbidden dark matter annihilation, the main focus of this work, which contains fermionic DM and a light singlet scalar that connects the DM and standard model (SM) leptons. 
The strong suppression of the scattering between DM and SM particles happens quite early in the DM depletion history, where the DM temperature drops away from the thermal equilibrium, $T_\chi < T_{\rm SM}$,  leading to the decreased kinetic energy of DM. 
The forbidden annihilation thus becomes inefficient since small kinetic energy cannot help exceed the annihilation threshold, naturally leading to a larger abundance.  To show the eKD discrepancy, we numerically solve the coupled Boltzmann equations that govern the evolution of DM number density and temperature. It is found that eKD significantly affects the DM abundance, resulting in almost an order of magnitude higher than that by the traditional calculation. We also discuss the constraints from experimental searches on the model parameters, where the viable parameter space shrinks when considering the eKD effect.
\end{abstract}

\maketitle

\section{Introduction}
\label{sec:intro}

Relic density is an essential topic for dark matter (DM) physics. The classical scenario that explains the observed abundance in the present Universe is thermal particle production in the early Universe, which is the so-called freeze-out mechanism for the weakly interacting massive particle (WIMP) \cite{Lee:1977ua, Hut:1977zn, Sato:1977ye, Dicus:1977nn, Wolfram:1978gp}. 
The DM particles were initially in thermal equilibrium with the heat bath via the intense interactions among them. 
The number density dilutes along with the expansion of the Universe and 
finally freezes out of the heat bath once the annihilation rate falls behind the cosmic expansion rate, 
resulting in a comoving constant.

Gondolo {\it et al.} \cite{Gondolo:1990dk,Edsjo:1997bg} have developed the renowned treatment of 
calculating the DM relic density by solving the Boltzmann equation of the number density with high accuracy, called the ``standard'' method. 
One hypothesis entering this treatment is that DM keeps in local kinetic equilibrium with 
the thermal plasma during or even after the freeze-out process. 
The scatterings with the standard model (SM) particles have been at a much more intense level \cite{Bringmann:2009vf}.

However, this is not always true for many well-motivated mechanisms, 
where kinetic equilibrium might decouple earlier than assumed, leading to the early kinetic decoupling (eKD) around the freeze-out period. 
The eKD effect has been extensively studied in the literature \cite{vandenAarssen:2012ag,Binder:2017rgn,Brummer:2019inq,Ala-Mattinen:2019mpa,Abe:2020obo,Binder:2021bmg,Zhu:2021vlz,Hryczuk:2021qtz,Abe:2021jcz,Du:2021jcj,Ala-Mattinen:2022nuj,Hryczuk:2022gay}. 
The most influenced scenarios include 
1) resonant annihilation of dark matter \cite{Binder:2017rgn,Abe:2020obo,Binder:2021bmg},  
2) Sommerfeld-enhanced annihilation \cite{Binder:2021bmg}, 
and 3) subthreshold annihilation  (also known as the forbidden annihilation) \cite{Binder:2021bmg}. 
In these regimes, the eKD occurs because the elastic scattering processes are suppressed; DM and the SM particles experience different temperatures.  
All these cases demonstrated the actual DM abundance can be affected by up to 1 order of magnitude compared to the traditional method, at least in some parts of the parameter space.

Here, we study the eKD in the forbidden DM scenario \cite{DAgnolo:2015ujb}. In such a scenario, DM dominantly annihilates into heavier final states, which can proceed at finite temperatures in the early Universe, relying on the thermal tail with the high velocity of DM. Many works were devoted to studying the forbidden annihilations with different theoretical models and a variety of phenomenological topics \cite{Delgado:2016umt,DAgnolo:2020mpt,Wojcik:2021xki,Yang:2022zlh,Cheng:2022esn,Herms:2022nhd,Herms:2022rou}. 
In this work, we employ the model where forbidden DM annihilations into the SM leptons are mediated by a singlet scalar. Such channels were studied in Ref.\cite{DAgnolo:2020mpt}, which is experimentally viable and predicts a very narrow mass range for DM, that can be tested at future beam dump experiments. However, the eKD effect has not been studied in the context of this model.

Actually, the scatterings of DM against SM particles are strongly suppressed in the early times because of the mass splitting between the DM and SM leptons. The DM temperature drops away from the thermal bath temperature, $T_\chi < T_{\rm SM}$, which leads to the decreased kinetic energy of DM. The forbidden annihilations thus become inefficient 
since small kinetic energy cannot help exceed the annihilation threshold. 
So this naturally leads to a larger abundance and the eKD effects in  such a model should not be neglected. 
On the contrary, eKD will cause significant impacts on DM abundance. 
We investigate the relic density beyond the standard treatment used in Ref. \cite{DAgnolo:2020mpt} by considering the coupled Boltzmann differential equations, 
where the temperature evolution of the dark sector could be taken into account. 
We use the public code DRAKE \cite{Binder:2021bmg} to perform the numerical calculations.
We find a DM relic density that differs by up to an order of magnitude from the standard treatment and leaves a reduced feasible parameter space under the various experimental constraints.

The rest of this paper is structured as follows.  
In Sec.  \ref{sec:eKD},
we start with a general description of the coupled Boltzmann equations that govern 
the evolution of DM number density and  temperature. 
In addition, we discuss the DM model for forbidden annihilations and analyze the occurrence of early kinetic decoupling. 
Section \ref{sec:relic} is devoted to a thorough study of DM relic density for the forbidden channels, and makes a detailed comparison between the new treatment and the traditional one. 
We further discuss various constraints from collider searches and astrophysical observations on the parameter space in Sec. \ref{sec:constraints}.
Finally, we conclude in Sec. \ref{sec:conclusion}.


\section{Early kinetic decoupling effects on forbidden annihilations}
\label{sec:eKD}

\subsection{Basic formulas}

Keeping kinetic equilibrium during and even after the freeze-out epoch is one underlying assumption for traditional relic density calculations. However, this is not always the case for various scenarios, where kinetic decoupling happens earlier than the chemical decoupling process. 
To study the DM relic density by taking into account the early kinetic decoupling effect, 
we should consider the following Boltzmann equation for DM phase-space distribution ~\cite{Binder:2017rgn,Binder:2021bmg} 
\begin{align}
 E \left( \frac{\partial}{\partial t} - H \vec{p}\cdot \frac{\partial}{\partial \vec{p}} \right) f_\chi (t, \vec{p})=
C_{\rm ann.}[f_\chi] + C_{\rm el.}[f_\chi],
\label{Eq:boltzmann}
\end{align}
where $E$ is the energy of the DM, $H$ is the Hubble constant, $\vec{p}$ is the momentum of DM,
and $f_\chi$ is the DM phase-space density.
The collision term $C_{\rm ann.}$ represents the annihilation of DM particles into thermal bath particles, and $C_{\rm el.}$ 
is for elastic scattering processes between DM and SM scattering partners. For two-body processes, 

\begin{eqnarray}
 C_{\rm ann.}&=&\frac{1}{2 g_\chi}
\sum
\int \frac{d^3 p'}{(2\pi)^3 2 E_{p'}}
\int \frac{d^3 k}{(2\pi)^3 2 E_k}
\int \frac{d^3 k'}{(2\pi)^3 2 E_{k'}}\nonumber\\
&\times&
(2\pi)^4 \delta^4(p + p' - k - k') \\
&\times& 
\Bigl(-|{\cal M}_{\chi \chi \to {\cal B} {\cal B}'}|^2 f_\chi(\vec{p}) f_\chi(\vec{p'}) (1 \pm f^{eq}_{\cal B}(\vec{k})) (1 \pm f^{eq}_{\cal B'}(\vec{k'}))\nonumber\\
&+&|{\cal M}_{{\cal B}{\cal B'} \to \chi \chi}|^2 f^{eq}_{\cal B}(\vec{k}) f^{eq}_{{\cal B}'}(\vec{k'}) (1 \pm f_\chi(\vec{p})) (1 \pm f_\chi(\vec{p'}))
\Bigr),\nonumber
\end{eqnarray}
\begin{eqnarray}
 C_{\rm el.}
&=&
\frac{1}{2g_\chi}
\sum
\int \frac{d^3 p'}{(2\pi)^3 2 E_{p'}}
\int \frac{d^3 k}{(2\pi)^3 2 E_k}
\int \frac{d^3 k'}{(2\pi)^3 2 E_{k'}}\nonumber\\
&\times&
(2\pi)^4 \delta^4(p + p' - k - k') \\
&\times& \Bigl(
-|{\cal M}_{\chi {\cal B} \to \chi {\cal B}}|^2 
f_\chi(\vec{p}) f^{eq}_{\cal B}(\vec{k}) (1 \pm f_\chi(\vec{p'})) (1 \pm f^{eq.}_{\cal B}(\vec{k'}))
\nonumber\\
&+&|{\cal M}_{\chi {\cal B} \to \chi {\cal B}}|^2 
f_\chi(\vec{p'}) f^{eq.}_{\cal B}(\vec{k'}) (1 \pm f_\chi(\vec{p})) (1 \pm f^{eq.}_{\cal B}(\vec{k}))
\Bigr),\nonumber
\end{eqnarray}
where ${\cal B}$ and ${\cal B}'$ stand for particles in the thermal bath such as SM leptons, 
$g_\chi$ is the number of internal degrees of freedom of DM,
and $f_{\cal B}^{eq}$ is given by the Fermi-Dirac or Bose-Einstein distribution depending on the spin of ${\cal B}$.
The summation should be taken for all the internal degrees of freedom
for all the particles.
For the nonrelativistic DM, $C_{\rm el.}$ can be simplified as the Fokker-Planck operator~\cite{Binder:2016pnr,Bertschinger:2006nq, Bringmann:2006mu, Bringmann:2009vf}
\footnote{As pointed out by Ref.\cite{Binder:2021bmg}, for forbidden DM, the Fokker-Planck approximation is not that accurate, but the dominant eKD effect on the relic density can in many cases still be fairly well captured by the Fokker-Planck approximation. For more precise treatment, we leave it to future work.}: 
\begin{align}
\label{Eq:FP}
C_\mathrm{el} 
&\simeq
\frac{E}{2} \gamma(T)
{\Bigg [}
T E\partial_p^2 \!+ \left(2 T  \frac{E}{p} \!+\! p \!+\! T\frac{p}{E}\right) \partial_p + 3
{\Bigg ]}f_{\chi}\,.
\end{align}
In the above, the momentum transfer rate $\gamma(T)$ is given by (see also 
Ref.~\cite{Gondolo:2012vh}) 
\begin{eqnarray}
\label{eq:gamma}
\gamma= \frac{1}{3 g_{\chi} m_{\chi} T} \! \int \! \frac{\text{d}^3 k}{(2\pi)^3} f_{\cal B}^{\pm}(E_k)\left[1\!\mp\! f_{\cal B}^{\pm}(E_k)\right] \! \! \! \int\limits^0_{-4 k_\mathrm{cm}^2} \! \! \!  \text{d}t (-t) \frac{\text{d}\sigma}{\text{d}t} v\,,
\end{eqnarray}
where the  differential cross section can be expressed as $({\text{d}\sigma}/{\text{d}t}) v$ $ \equiv$ $|\mathcal{M}|^2_{\chi f\leftrightarrow\chi f}$ /$(64 \pi {k} E_k m_\chi^2)$, and 
$k_{\rm cm}^2$ is given by 
\begin{eqnarray}
 k_{\rm cm}^2 =\frac{m_\chi^2 ( E_k^2 - m_{\cal B}^2)}{m_\chi^2 + m_{\cal B}^2 + 2 m_\chi E_k}.
\end{eqnarray}
Here $E_k$ is the energy of  heat bath particle ${\cal B}$.
Note that $k_{\rm cm}^2 \neq  E_k^2 - m_{\cal B}^2 = |\vec{k}|^2$.


During the chemical decoupling, the scattering processes may not be frequent enough to maintain the kinetic equilibrium, which means that DM particles own a different temperature $T_\chi$ from the thermal plasma in their following evolution.  
A common definition of the DM temperature is 
\begin{eqnarray}
 T_\chi
&=&
\frac{g_\chi}{3 n_\chi}\int \frac{d^3 p}{(2\pi)^3} \frac{\vec{p}^2}{E} f_\chi (\vec{p})\equiv
\frac{s^{2/3}}{m_\chi} y,
\end{eqnarray}
which is also a function of the thermal bath temperature $T$. In this definition $n_\chi$ is the number density of the DM,
and $s$ is the entropy density.  Here $y$ is a dimensionless version in analogy to the DM yield $Y (=n_\chi/s)$.

To reach a suitable description of the DM temperature evolution and then explore the eKD effect on the chemical decoupling process, we should 
consider the second moment of  $f_\chi$ as a dynamical degree of freedom. 
By integrating Eq.~(\ref{Eq:boltzmann}) with $g_\chi \int \frac{d^3 p}{(2\pi)^3} \frac{1}{E}$
and
$g_\chi \int \frac{d^3 p}{(2\pi)^3} \frac{1}{E} \frac{\vec{p}^2}{E^2}$, one obtains the zeroth and second moments of the Boltzmann equation, respectively. 
This leads to a relatively simple coupled system of Boltzmann differential equations ( denoted as the \textbf{cBE} method hereafter), 
\begin{eqnarray}
\frac{Y'}{Y} &=& \frac{s Y}{x \tilde H}\left[
\frac{Y_{\rm eq}^2}{Y^2} \left\langle \sigma v\right\rangle_T- \left\langle \sigma v\right\rangle_{T_\chi}
\right]\,, \label{Yfinalfinal}\\
\frac{y'}{y} &=&   \frac{1}{x\tilde H}\langle C_{\text{el}} \rangle_2
+\frac{sY}{x\tilde H}\left[
\left\langle \sigma v\right\rangle_{T_\chi}-\left\langle \sigma v\right\rangle_{2,T_\chi}
\right] \label{yfinalfinal}\\ 
&&+\frac{sY}{x\tilde H}\frac{Y_{\rm eq}^2}{Y^2}\left[
\frac{y_{{\rm eq}}}{y}\left\langle \sigma v\right\rangle_{2,T}\!-\!\left\langle \sigma v\right\rangle_T
\right]
+2(1-w)\frac{H}{x\tilde H}\,, \nonumber
\end{eqnarray}
where $x$ is defined as usual $x = m_\chi/T$ and 
$Y_{\rm eq}(x)\equiv n_{\rm eq}(T)/s$. 
$\tilde H\equiv H/\left[1+ (1/3)d(\log g^s_{\rm eff})/d(\log T)\right]$, 
with $g^s_{\rm eff}$ being the entropy degrees of freedom of the 
background plasma. 
$w(T_\chi)\equiv 1-{\langle p^4/E^3 \rangle_{T_\chi}}/({6T_\chi})$, with $\langle p^4/E^3\rangle
=\frac{g_\chi}{n_\chi^{eq}(T_\chi)}\int \frac{d^3p}{(2\pi)^3} {\left( \vec{p} \cdot \vec{p} \right)^2}/{E^3} e^{- \frac{E}{T_\chi}}$. 
Note that the elastic scattering term given in Eq.~(\ref{Eq:FP}) does not contribute to 
the zeroth moment term. This is a natural consequence because the elastic scattering processes do not change the number density of DM.

The above compact form of the differential equations contains the following thermally averaged cross sections, 
\begin{align}
\langle C_{\text{el}} \rangle_2 \equiv \frac{g_\chi}{3 n T_{\chi}} \int \frac{\text{d}^3 p}{(2\pi)^3} \frac{p^2}{E^2} C_{\text{el}}\;,
\end{align}
\begin{eqnarray}
\VEV{\sigma v}_{T_\chi}
&\equiv&
\frac{g_\chi^2}{(n_\chi^{eq})^2}
\int \frac{d^3 p}{(2\pi)^3} 
\int \frac{d^3 q}{(2\pi)^3} 
\left( \sigma v \right)_{\chi \chi \to {\cal B} {\cal B}'}\nonumber\\
&\times&
f_\chi^{eq}(\vec{p}, T_\chi)
f_\chi^{eq}(\vec{q}, T_\chi)
,
\end{eqnarray}
The thermal average $\langle \sigma v\rangle_{2,T}$ is a variant of 
the commonly used thermal average $\langle \sigma v\rangle_T$, and is explicitly stated in 
Ref.~\cite{Binder:2017rgn} and introduced as 
\begin{eqnarray}
\VEV{\sigma v}_{2, T_\chi}
&=&
\frac{g_\chi^2}{(n_\chi^{eq})^2 T_\chi}
\int \frac{d^3 p}{(2\pi)^3} 
\int \frac{d^3 q}{(2\pi)^3} 
\frac{\vec{p} \cdot \vec{p} }{3E} 
\left( \sigma v \right)_{\chi \chi \to {\cal B} {\cal B}'} \nonumber\\
&\times&
f_\chi^{eq}(\vec{p}, T_\chi)
f_\chi^{eq}(\vec{q}, T_\chi). 
\end{eqnarray}

For 
$\VEV{\sigma v}_{T} $
and 
$\VEV{\sigma v}_{2, T} $,
replace $T_\chi$ by $T$ in 
$\VEV{\sigma v}_{T_\chi} $
and 
$\VEV{\sigma v}_{2, T_\chi} $, respectively.
And $n_\chi^{eq}(T_\chi)$ is given by
\begin{eqnarray}
 n_\chi^{eq}(T_\chi) = g_\chi \int \frac{d^3 p}{(2\pi)^3} f_\chi^{eq}(\vec{p}, T_\chi)
= g_\chi \int \frac{d^3 p}{(2\pi)^3} e^{- \frac{E_p}{T_\chi}}.
\end{eqnarray} 

In this work, we use the numerical routine DRAKE to solve the coupled Boltzmann equations. 
The measured value of $\Omega h^2$ by the Planck Collaboration
is $\Omega h^2 = 0.120\pm 0.001$~\cite{Planck:2018vyg}. 
The viable parameter space is determined by matching this value. 

\subsection{Model and discussion on forbidden channels}

We have adopted a simple model that only takes into account DM annihilations into SM leptons. The DM is a Dirac fermion coupled to the SM sector via the scalar portal $\phi$. After the electroweak symmetry breaking, the effective Lagrangian can be written as
\begin{equation}
-\mathcal{L}\supset g_{ij} \phi \bar l_i l_j+g^{A}_{ij} \phi \bar l_i \gamma_5 l_j + g_{\chi} \phi \bar \chi \chi + g^A_\chi \phi \bar \chi \gamma_5 \chi,  
\label{eq:scalar}
\end{equation}
where the indices on the couplings $i,j= e, \mu, \tau$. This model has been studied thoroughly for the forbidden mechanism (for detail, refer to Ref. \cite{DAgnolo:2020mpt}). 
The merits include the following:  
1) DM mass is limited in quite a small window close to the masses of the SM leptons, which is a strong prediction that can be tested soon by colliders or beam-dump experiments. 
2) Kinematically forbidden DM naturally evade the stringent constraints from the energy injections into the cosmic microwave background (CMB) ~\cite{DAgnolo:2015ujb}. In the forbidden DM scenario, the energy injection processes suffer the Boltzmann suppression at $T\lesssim$~eV, so that sub-GeV thermal relics are consistent with the experiment, making annihilations to SM leptons with DM masses $m_\chi \ll 10$~GeV still viable.

In this scenario, the DM relic density should be carefully scrutinized, as 
the eKD effect appears generic. The actual relic density receives a significant correction compared with the conventional method, as shown in the following sections.

Following Refs.~\cite{DAgnolo:2015ujb, DAgnolo:2020mpt}, we consider a pair of DM particles dominantly annihilating into two SM particles $2 \ell$ with mass $m_\chi < m_{\ell}$. 
Cosmological constraints make forbidden annihilations into electrons unfeasible, including big bang nucleosynthesis (BBN) and CMB \cite{Sabti:2019mhn}. 
So, for simplicity, we only consider the $\mu^+\mu^-$ and $\tau^+\tau^-$ channels, with abbreviated couplings $g^{(A)}_\mu,~g^{(A)}_\tau,$ and $g_\chi^{(A)}$, which allows us to explore all the relevant DM phenomenology systematically. 

Applying the detailed balance condition for the DM number-changing process, the cross section of the forbidden channels is exponentially suppressed,  
\begin{equation}
\langle \sigma_\chi v \rangle= \langle \sigma_\ell v \rangle \frac{(n_{\ell}^{\rm eq})^2 }{(n_\chi^{\rm eq})^2} \simeq \langle \sigma_\ell v \rangle e^{- 2\Delta x}\, ,
\end{equation} 
where  $\Delta\equiv (m_{\ell}- m_\chi)/ m_\chi$. $\sigma_\chi \equiv \sigma(\chi \chi \to \ell \ell_)$, while $\sigma_\ell$ is the cross section for the inverse process. When the annihilation rate becomes slower than the Hubble expansion, DM is no longer in equilibrium with the SM thermal bath, resulting in chemical decoupling.

What about the scattering between DM and SM particles during this period? From Eq.~(\ref{eq:gamma}), the momentum transfer rate $\gamma$ is proportional to an exponential factor:  
\begin{equation}
\gamma (x) \propto e^{-(\Delta+1) x}. 
\label{eq:gamma-semi}
\end{equation} 
The full expression of the rates is listed in Appendix \ref{app:rate}. 
In the forbidden scenario, $\Delta+1>1$ implies the scattering frequency experiences a strong suppression at a much earlier period.

\begin{figure}[!htbp]
    \centering
    \includegraphics[width=0.45\textwidth]{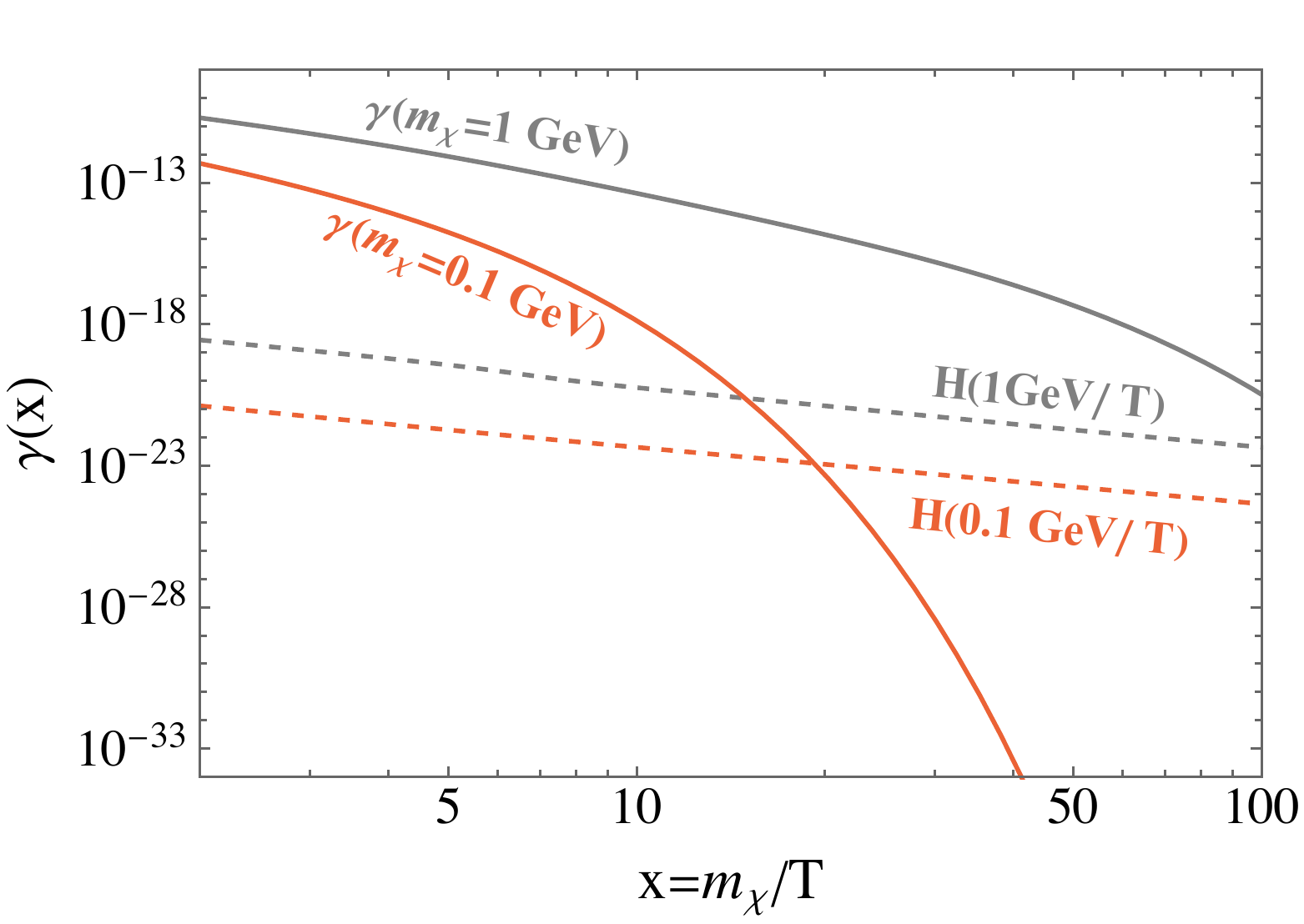}
    \caption{Evolution of momentum transfer rate $\gamma(x)$ for  $\chi \mu^\pm\to\chi \mu^\pm$ scattering, and comparing with Hubble constant H(x). The red lines stand for the forbidden scenario, where $m_\chi=0.1{\rm GeV}$ and $m_\phi=0.26 {\rm GeV}$. The gray lines correspond to the nonforbidden case, that we take as $m_\chi=1 {\rm GeV}$.}
    \label{fig:gamm}
\end{figure}

It is known that DM kinetically decouples out of the 
SM thermal bath as long as the momentum transfer rate $\gamma$ is smaller than the Hubble expansion, $\gamma<H$. 
For illustration, we show the momentum transfer rate $\gamma$ for the $\chi\mu^\pm\to\chi\mu^\pm$ scattering process, in Fig. \ref{fig:gamm}. In comparison, we plot the Hubble parameter $H(x)$ as a function of $x$. The red lines stand for the evolution of $\gamma (x)$ and $H(x)$ in the forbidden DM case where we take $m_\chi=0.1 {\rm GeV}$. To demonstrate the distinctiveness of the forbidden DM, we also provide the results of a nonforbidden case where DM mass is larger than that of the annihilation products ($m_\chi$=1 {\rm GeV}). 
The most remarkable finding is that $\gamma$ of the forbidden case becoming comparable with $H(x)$ happens much earlier than that of the nonforbidden case. The kinetic decoupling starts at around $x=20$, which is usually the same time as DM chemically decoupled from the thermal bath. 
The reason for this very early kinetic decoupling is straightforward to understand as the result of an exponential suppressed momentum transfer rate, as derived in Eq.~(\ref{eq:gamma-semi}).
One can obtain  similar results for the $\chi\bar\chi\to\tau^+\tau^-$ forbidden channel. We conclude that eKD exists in the forbidden DM scenario.

The next step is to find the effect of the eKD more concretely. We systematically study DM relic density in both {\bf cBE} and traditional methods (denoted as {\bf nBE} as in Ref.~\cite{Binder:2021bmg}) and then discuss the phenomenological possibilities.
\section{Relic density: comparison between {\bf cBE} and {\bf nBE} approaches}
\label{sec:relic}

In this section, we compute the relic density in both the standard method ({\bf nBE}) and the {\bf cBE} approach. 
We restrict our study to the same range of $m_\chi$ that $0.9 m_\mu \lesssim m_\chi \lesssim m_\mu$ as derived in Ref.\cite{DAgnolo:2020mpt}, in which forbidden annihilation into $\mu^+\mu^-$ is experimentally viable. 
For the $\chi\bar\chi\to\tau^+\tau^-$ case, the corresponding DM mass is $0.8 m_\tau \lesssim m_\chi \lesssim m_\tau$. 
The detailed annihilation cross sections and the momentum transfer rates for scatterings are presented in Appendix \ref{app:rate}.

\begin{table*}[!htbp]
  \centering
  \caption{Benchmark points, selected for $\chi\bar\chi\to\mu^+\mu^-$ and $\chi\bar\chi\to\tau^+\tau^-$ channels. We set $g_\chi=0$ and $g^A_\chi=1.121$ for all cases as in Ref.\cite{DAgnolo:2020mpt} which can make a direct comparison.}
  \label{Tab:benchmarks}
  \begin{tabular}{ccccccccccccc}
  \toprule[1pt]
  \midrule[1pt]
  Benchmark ($\mu^+\mu^-$) & $\quad$ & $m_\chi$ & $\quad$ & $m_\phi$ & $\quad$ & $g_\mu$ & $\quad$ & $g_\mu^A$ &  $\quad$ & {$\Omega_{\rm cBE}h^2/\Omega_{\rm nBE}h^2$}\\
  \midrule[1pt]
  BP$\mu$1 & $\quad$ & 0.1 GeV & $\quad$ & 0.26 GeV & $\quad$ & 0.00343326 &  $\quad$ & 0  & $\quad$ & 0.12/0.018\\
  BP$\mu$2 & $\quad$ & 0.1 GeV & $\quad$ & 0.26 GeV & $\quad$ & 0.00091757 & $\quad$ & $g_\mu$  & $\quad$ & 0.12/0.0186\\
  BP$\mu$3 & $\quad$ & 0.1055 GeV & $\quad$ & 0.3 GeV & $\quad$ & 0.00145082 &  $\quad$ & 0 & $\quad$ & 0.12/0.036\\
  BP$\mu$4 & $\quad$ & 0.1055 GeV & $\quad$ & 0.3 GeV & $\quad$ & 0.00026751 & $\quad$ &  $g_\mu$   & $\quad$ & 0.12/0.064 \\
\midrule[1pt]
Benchmark ($\tau^+\tau^-$) & $\quad$ & $m_\chi$ & $\quad$ & $m_\phi$ & $\quad$ & $g_\tau$ & $\quad$ & $g_\tau^A$  & $\quad$ & {$\Omega_{\rm cBE}h^2/\Omega_{\rm nBE}h^2$}\\
  \midrule[1pt]
  BP$\tau$1 & $\quad$ & 1.6 GeV & $\quad$ & 5 GeV & $\quad$ & 1.09261 & $\quad$ & 0  & $\quad$ & 0.12/0.0088 \\
  BP$\tau$2 & $\quad$ &  1.6 GeV & $\quad$ & 5 GeV & $\quad$ & 0.268806 & $\quad$ & $g_\tau$ & $\quad$ & 0.12/0.0078\\
  BP$\tau$3 & $\quad$ & 1.77 GeV & $\quad$ & 6 GeV & $\quad$ & 0.0565632 & $\quad$ & 0 & $\quad$ & 0.12/0.032\\
  BP$\tau$4 & $\quad$ & 1.77 GeV & $\quad$ & 6 GeV & $\quad$ & 0.0096236 & $\quad$ & $g_\tau$  & $\quad$ & 0.12/0.058 \\
\midrule[1pt]
\bottomrule[1pt]
\end{tabular}
\end{table*}

To reveal the effects of the eKD and the differences between the cBE and nBE approaches, 
we first find several benchmark points for the two forbidden channels, 
shown in Table~\ref{Tab:benchmarks}. 
We also set $g_\chi=0,~(g^A_\chi)^2/4\pi=0.1$ as in Ref.\cite{DAgnolo:2020mpt}, for making a rough but straight comparison. The values of the rest model parameters are fixed to obtain the observed DM relic density for $\Omega_{\rm cBE}h^2$. With these inputs, we can compute the relic density in the {\bf nBE} method. It can be seen that the ratio of $\Omega_{\rm cBE}/\Omega_{\rm nBE}$ is sizable, even reaching an order of magnitude. 

\begin{figure}[!htbp]
    \centering
    \includegraphics[width=0.228\textwidth]{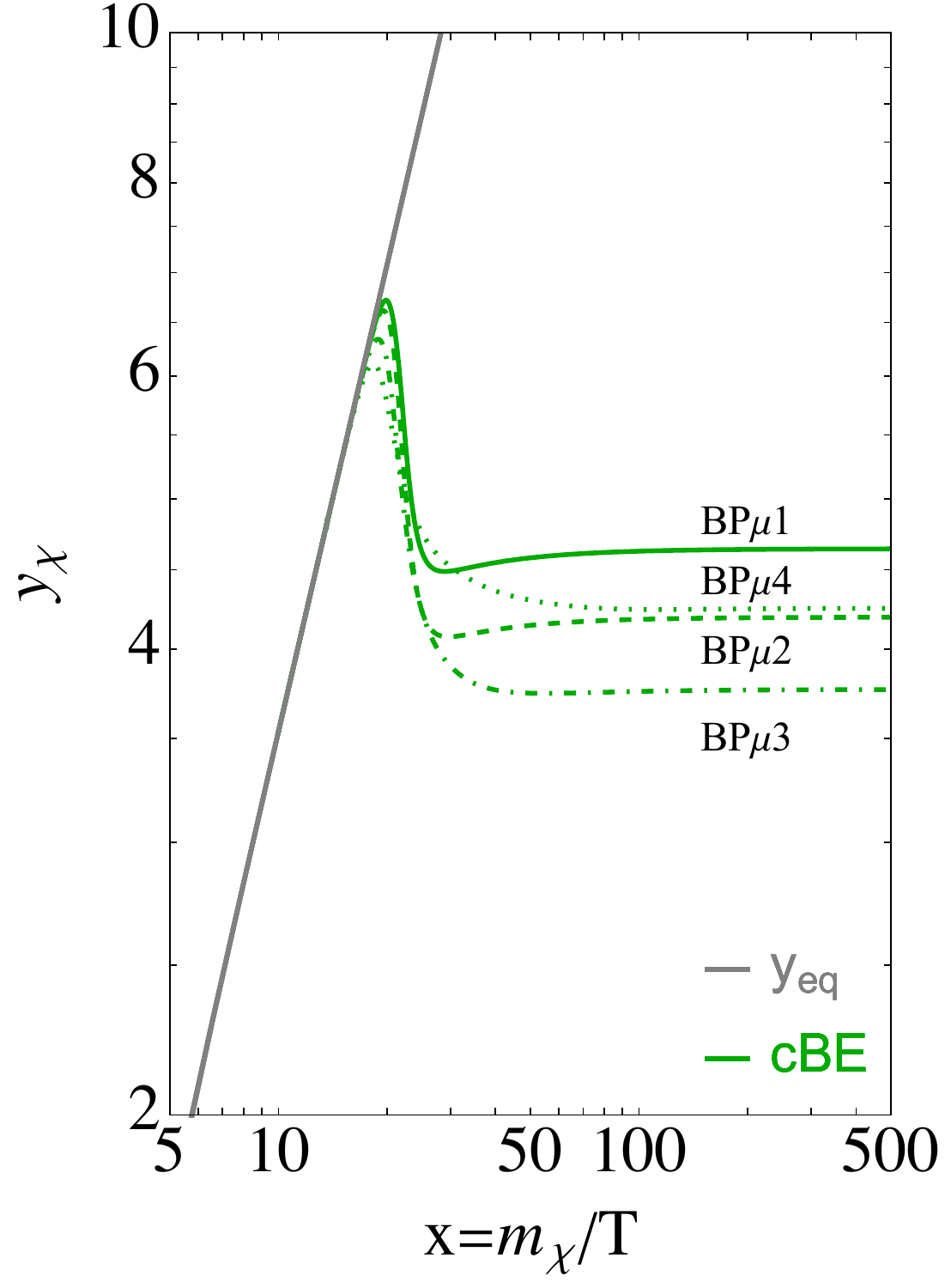}
    \includegraphics[width=0.248\textwidth]{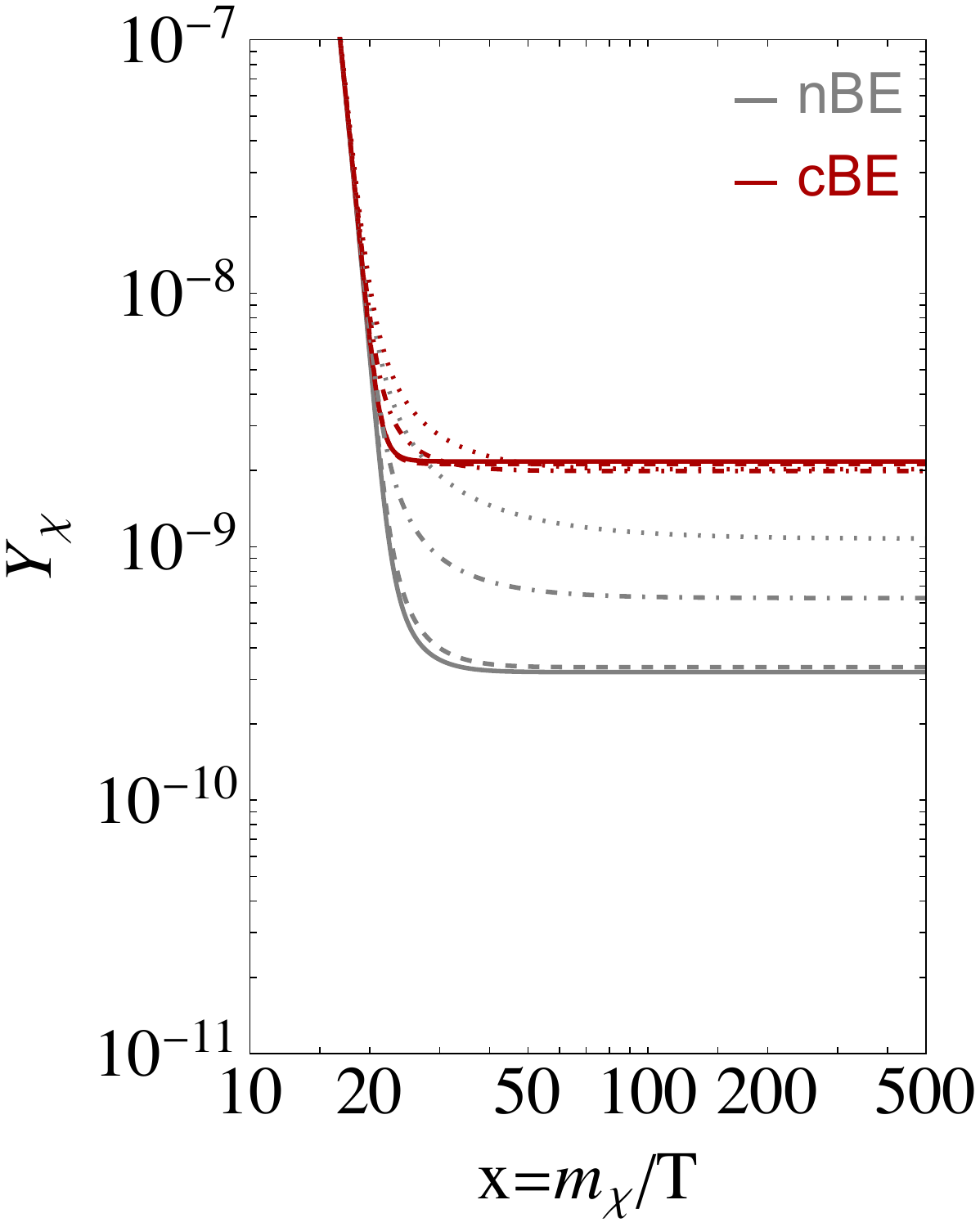}
     \caption{Evolution of DM temperature $y$ and  DM abundance $Y$. Left: the green lines depict the DM temperature evolution in the {\bf cBE} approach, with benchmarks BP$\mu$1 (solid), BP$\mu$2 (dashed), BP$\mu$3 (dot-dashed), and BP$\mu$4 (dotted). The gray line stands for the temperature of the thermal bath. Right: the red lines are the DM yields in the {\bf cBE} approach, for different parameter settings. The gray lines represent the yields in the traditional {\bf nBE} approach. The convention of the line styles is the same as in the left diagram.}
    \label{fig:Yy-x}
\end{figure}

The significant difference between the {\bf cBE} and {\bf nBE} results exactly comes from the eKD effects. 
In Fig.~\ref{fig:Yy-x}, we show the temperature and abundance evolution for selected benchmarks in Tab. \ref{Tab:benchmarks}. 
From the left panel, the green lines are the evolution curves of $y_\chi$, namely the temperature of DM, which depart from the thermal bath temperature (the gray line) at around $x=20$. 
Qualitatively, DM needs higher momenta to overcome the annihilation threshold, leading to a self-cooling phase as soon as it is no longer kinetically coupled to the muons. This is why there is a drop and the temperatures evolve separately for the dark sector and the SM sector since then. 
DM annihilation becomes less efficient much earlier just because of this cooling, which results in a higher DM abundance than in the {\bf nBE} approach, as shown in the right panel of Fig. \ref{fig:Yy-x}. 
Note that the same cooling phenomena also have been found in Ref. \cite{Binder:2021bmg}.


\begin{figure*}[!htbp]
  \includegraphics[width=0.45\textwidth]{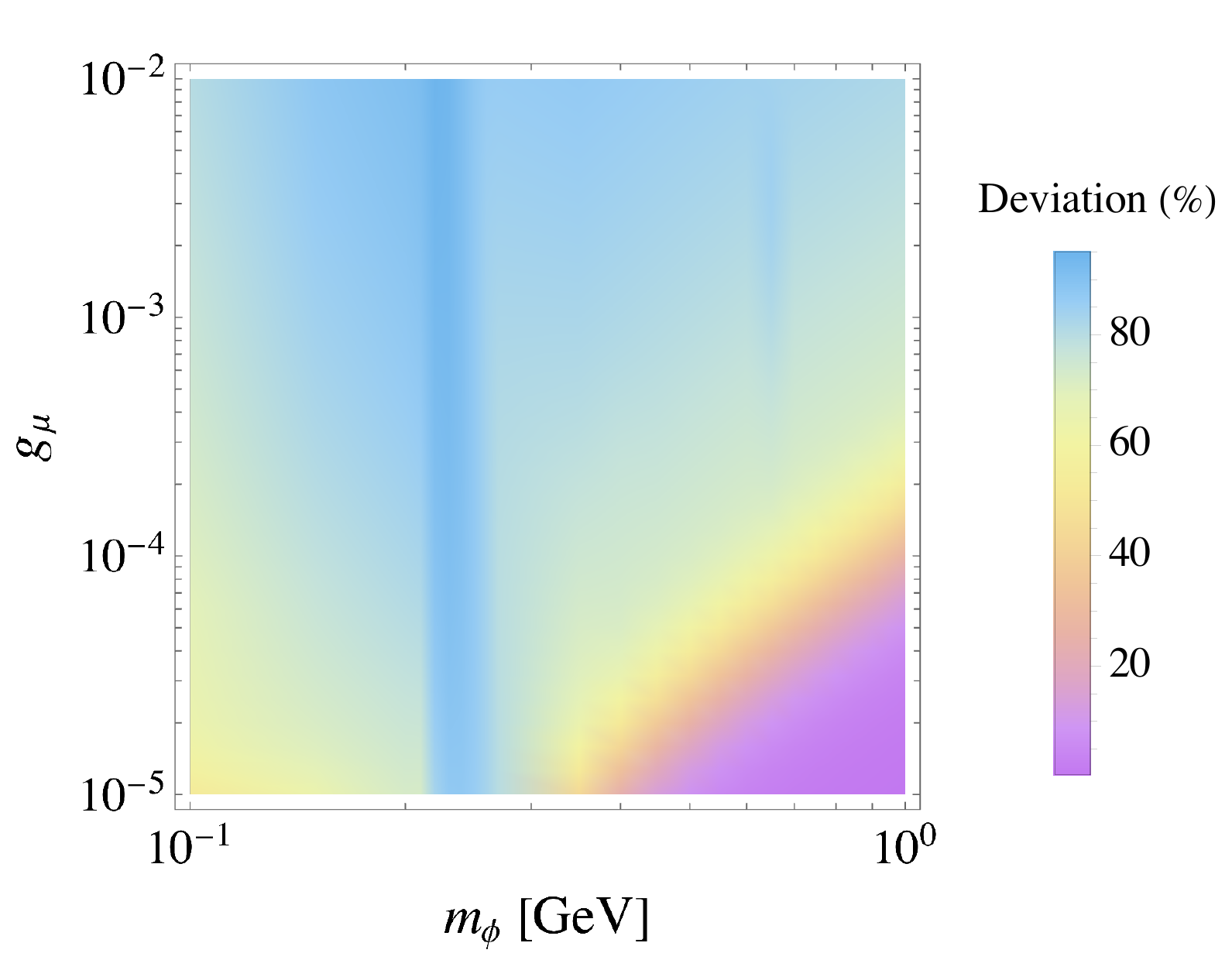} \qquad
  \includegraphics[width=0.443\textwidth]{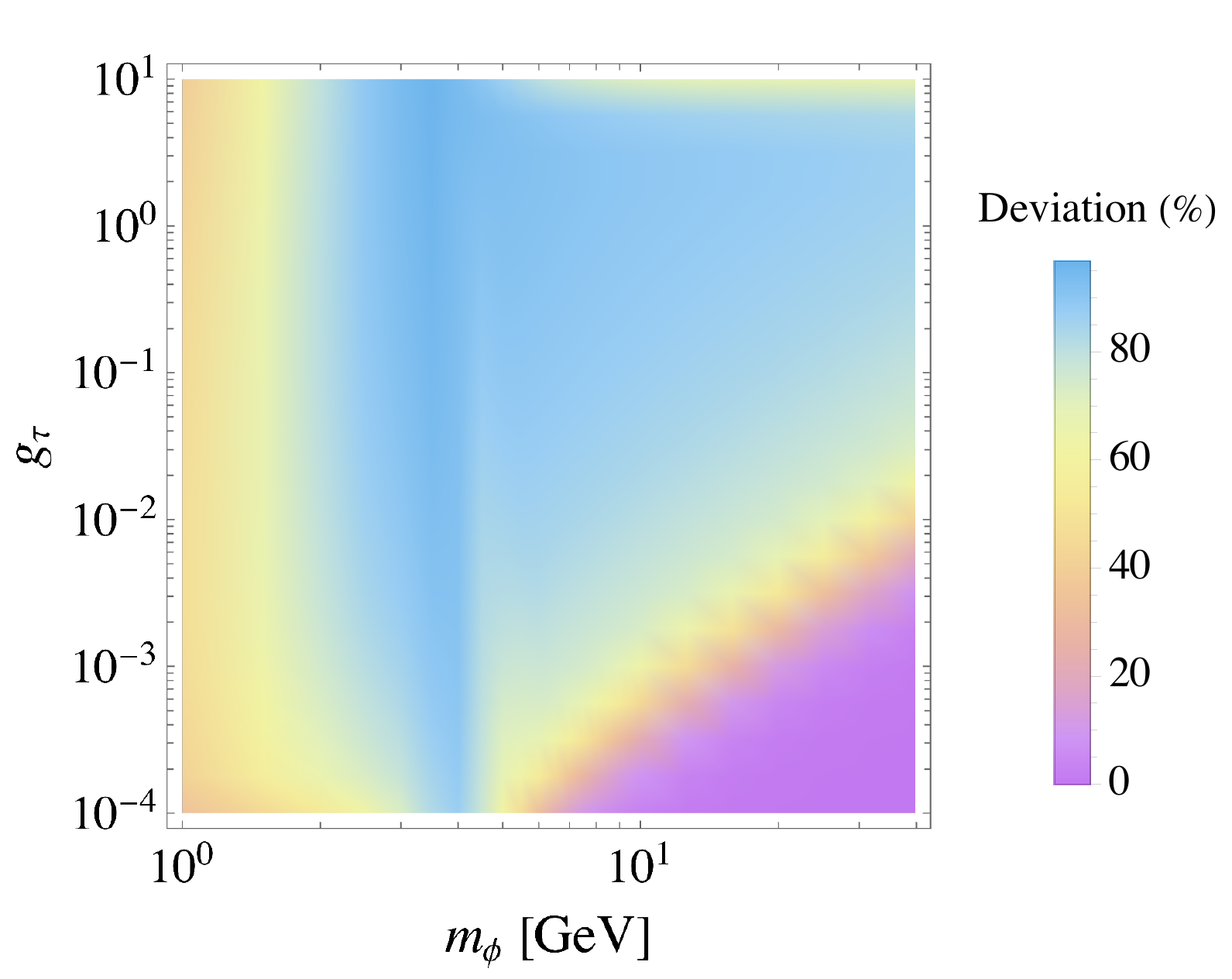}
   \caption{Density distribution of the deviation between {\bf cBE} and {\bf nBE} approaches, which is defined as $(\Omega_{\rm cBE}-\Omega_{\rm nBE})/\Omega_{\rm cBE}$. The left panel is for the $\chi\bar{\chi}\to\mu^+\mu^-$ channel and the right is for the  $\chi\bar\chi\to\tau^+\tau^-$ channel; we take $m_\chi=0.1$ GeV and $m_\chi=1.48$ GeV for the left and right panels respectively.}
  \label{Fig:deviation}
\end{figure*}

In Fig. \ref{Fig:deviation}, we show a global picture of the eKD effect for the forbidden cases of $\chi\bar\chi\to\mu^+\mu^-$ and $\chi\bar\chi\to\tau^+\tau^-$, where we define the deviations of the relic density in {\bf cBE} and {\bf nBE} approaches by ${\rm Deviation}\equiv(\Omega_{\rm cBE}-\Omega_{\rm nBE})/\Omega_{\rm cBE}$. 
We display the results in the $(m_\phi,~g_{\mu/\tau})$ plane by the density plotting method, with setting $m_\chi=0.1 {\rm GeV}$ and $m_\chi=1.48 {\rm GeV}$ for illustration. 
The sizeable deviations appear in most parameter spaces from 20\% up to almost 100\%. 
The maximum deviations seem to emerge in the resonance region. However, it should be noted that eKD effects already exist in the resonant annihilations of DM  \cite{Binder:2017rgn, Abe:2020obo, Binder:2021bmg}. So in this region, one should study the eKD for resonance and forbidden channels together. It is beyond the scope of our work, as we mainly focus on the forbidden annihilations. 

\begin{figure*}[!htbp]
  \includegraphics[width=0.445\textwidth]{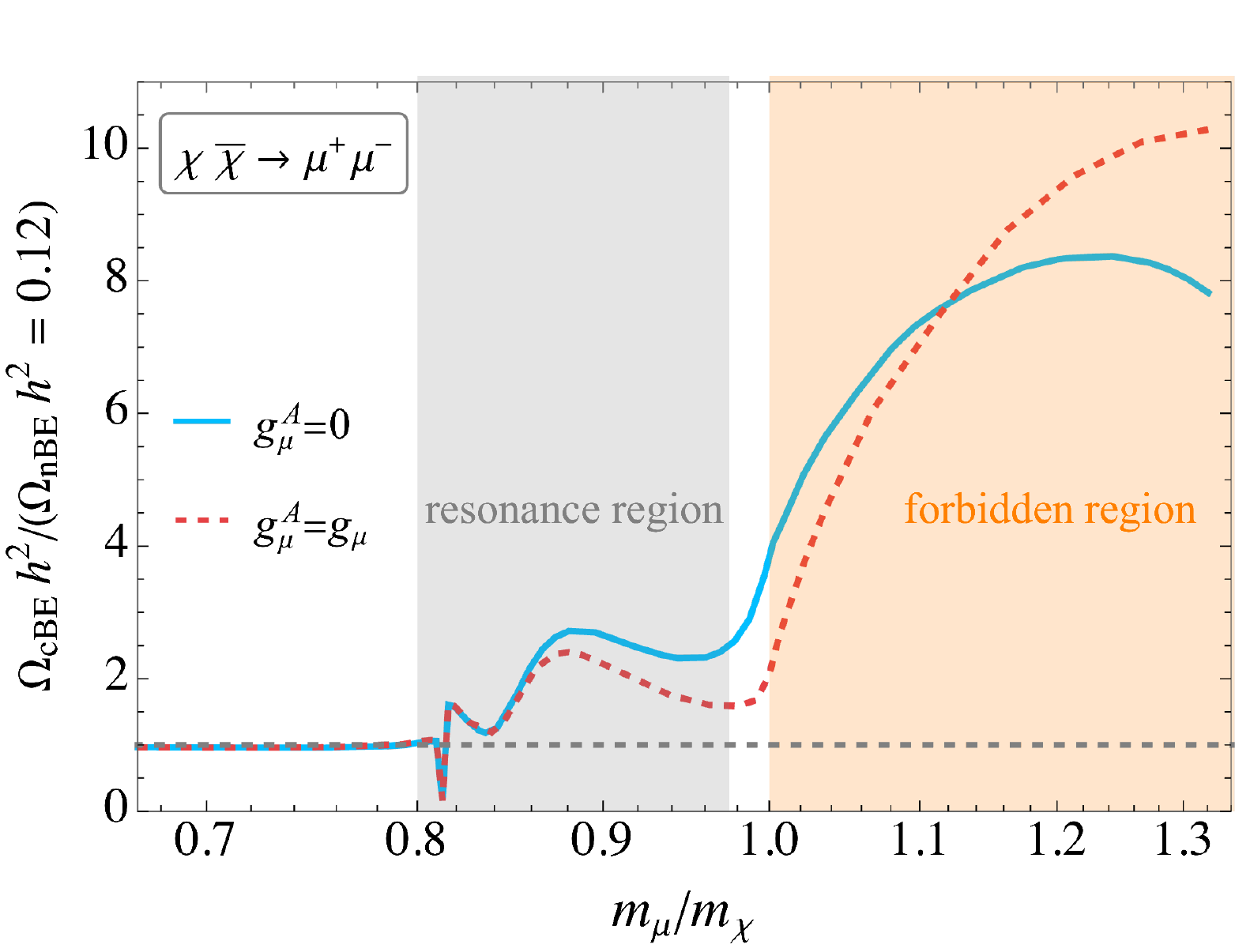}\qquad
  \includegraphics[width=0.445\textwidth]{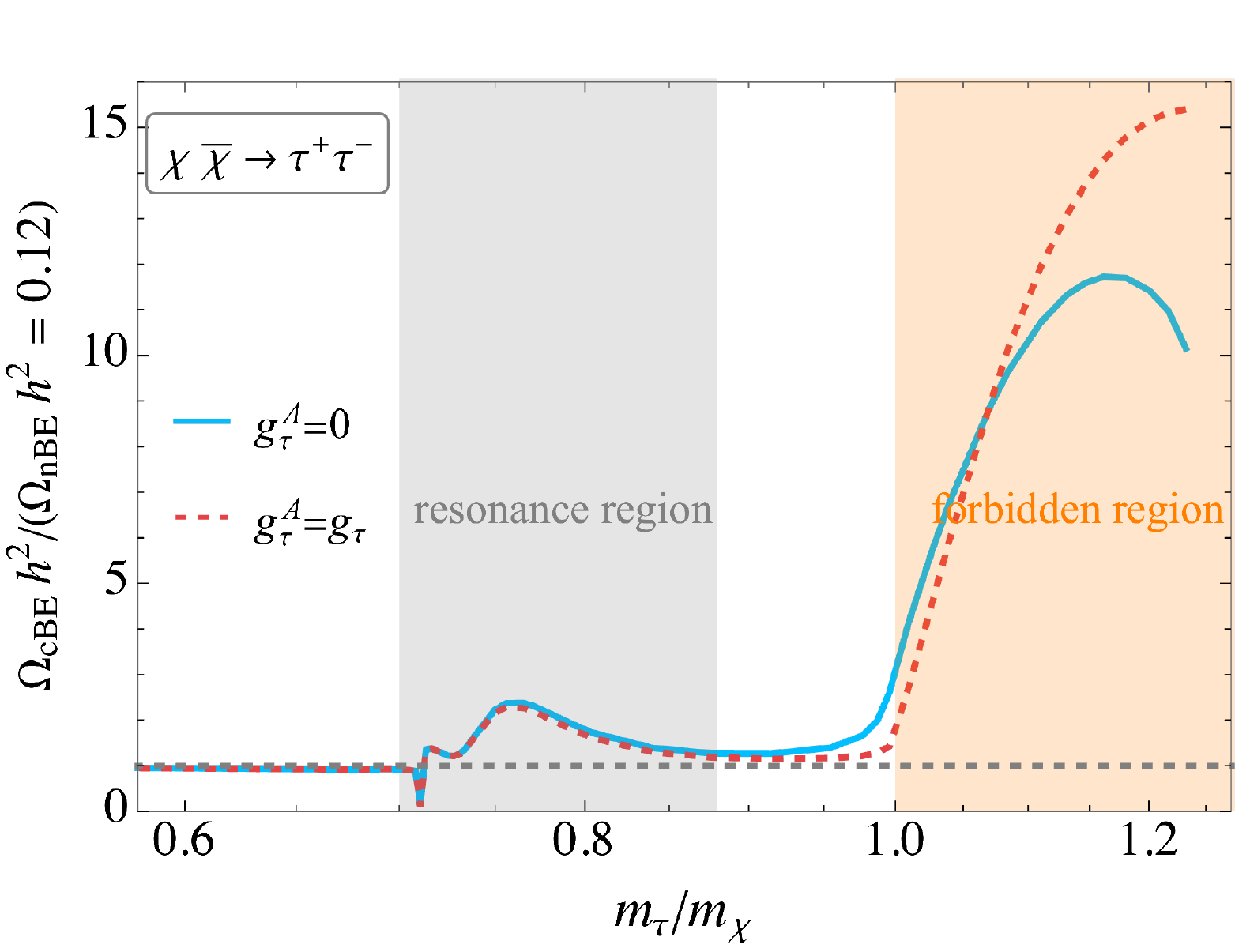}
  \caption{Relic density comparison between $\Omega_{\rm cBE} h^2$ and $\Omega_{\rm nBE} h^2$ for the $\chi\bar\chi\to\mu^+\mu^-$ and  $\chi\bar\chi\to\tau^+\tau^-$ channels. 
  Here, we choose the parameters to match the observed abundance in the {\bf nBE} approach. 
  The left diagram corresponds to the $\mu^+\mu^-$ mode, and we take $m_\phi=0.26$ GeV, and the right diagram presents the $\tau^+\tau^-$ mode, for which we take $m_\phi=5$ GeV.}
  \label{Fig:ratio}
\end{figure*}

To emphasize the importance of the improved treatment of the decoupling history near the mass threshold, we plot in Fig. \ref{Fig:ratio} the ratio of the resulting relic density to that of the standard {\bf nBE} approach. Here the parameters satisfy the requirement for $\Omega_{\rm nBE}  h^2=0.12$. 
The different choices of coupling $g^A_{l}$ correspond to different curves, as labeled in the plots. 
{The mass ratio $m_\ell/m_\chi$ can be divided into three regions. {At lower mass ratios, dark matter evolves as ordinary WIMPs where kinetic equilibrium is maintained.} The relic density derived by the {\bf cBE} and {\bf nBE} methods agrees with each other. 
{The gray-shaded region is known as the resonance region, where $2 m_\chi \simeq m_\phi$. Here, the eKD effect arises due to the distinct cooling and heating effects of dark matter.For a more detailed discussion of the origin of these features, please refer to Ref.\cite{Binder:2017rgn}.}
In the forbidden DM mass region (brown-shaded), we can see that the {\bf cBE} results are larger than {\bf nBE} several times with the same parameters. For $\mu^+\mu^-$ case, $3\lesssim\frac{\Omega_{\rm cBE}}{\Omega_{\rm nBE}}\lesssim10$, and $2\lesssim\frac{\Omega_{\rm cBE}}{\Omega_{\rm nBE}}\lesssim15$ for the $\tau^+\tau^-$ forbidden case. }

\begin{figure}[!htbp]
  \includegraphics[width=0.45\textwidth]{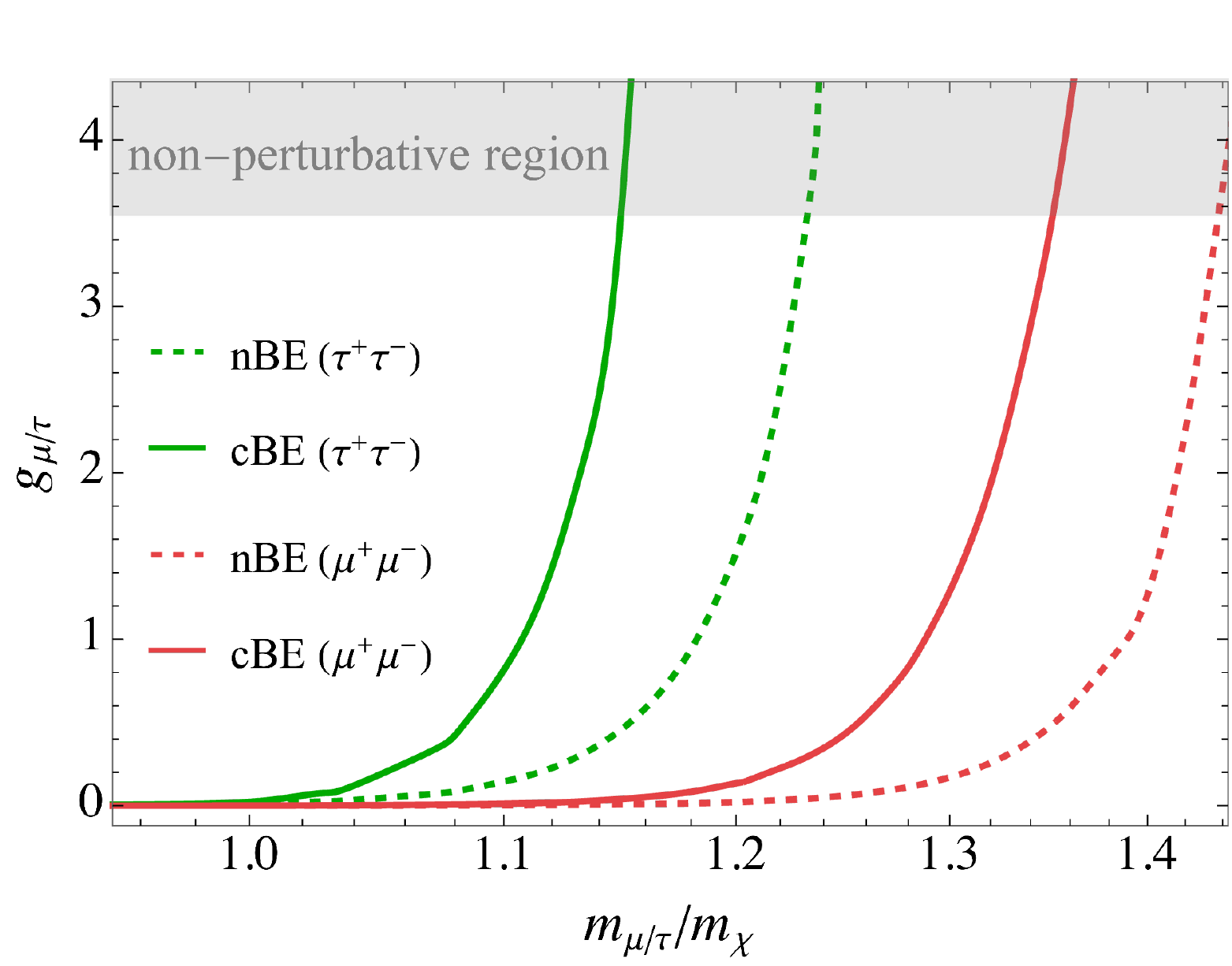}
  \caption{Correct relic density as function of mass ratio and coupling $g_{\mu/\tau}$ in both {\bf nBE} and {\bf cBE} approaches. 
  We take $m_\phi=0.26$ GeV for the $\mu^+\mu^-$ mode, 
  and $m_\phi=5$ GeV for the $\tau^+\tau^-$ mode.}
  \label{Fig:ratio-coupling}
\end{figure}

As the mass ratio $m_{\mu/\tau}/m_\chi$ increases, the coupling ($g_{\mu/\tau}$), which is needed to obtain the correct relic density, rises rapidly (see Fig. \ref{Fig:ratio-coupling}). 
We find the upper bounds of the mass ratio, beyond which the couplings become non-perturbative (i.e., $g_{\mu/\tau}<\sqrt{4\pi}$). Since a larger coupling is required in the {\bf cBE} approach, a smaller mass ratios are allowed compared to the standard  {\bf nBE} treatment.


Speaking overall, DM relic density in the {\bf cBE} method is larger than that by using the standard {\bf nBE} method. 
When the elastic scattering is strongly suppressed, {the temperature of DM particles drops below that of  the thermal bath ($T_\chi < T_{\rm SM}$),} meaning DM particles do not have enough kinetic energy. The forbidden annihilation thus becomes ineffective due to the inability to overcome the annihilation threshold, leading to a larger abundance in {\bf cBE} treatment.

\section{Parameter space and experimental constraints}
\label{sec:constraints}

In this section, we will find the feasible parameter space for the {\bf cBE} approach in the $(m_\phi,~g_{\mu,\tau})$ plane, by requiring the correct DM relic density. 
The viable parameter space further displays  more accurate results compared with {\bf nBE}.
Of course, there are numerous constraints on the model parameters that are imposed by the collider searches and the astrophysical observations across a wide range.

\subsection{$\chi \bar{\chi} \to \mu^+\mu^-$}

\begin{figure*}[!htbp]
  \includegraphics[width=0.475\textwidth]{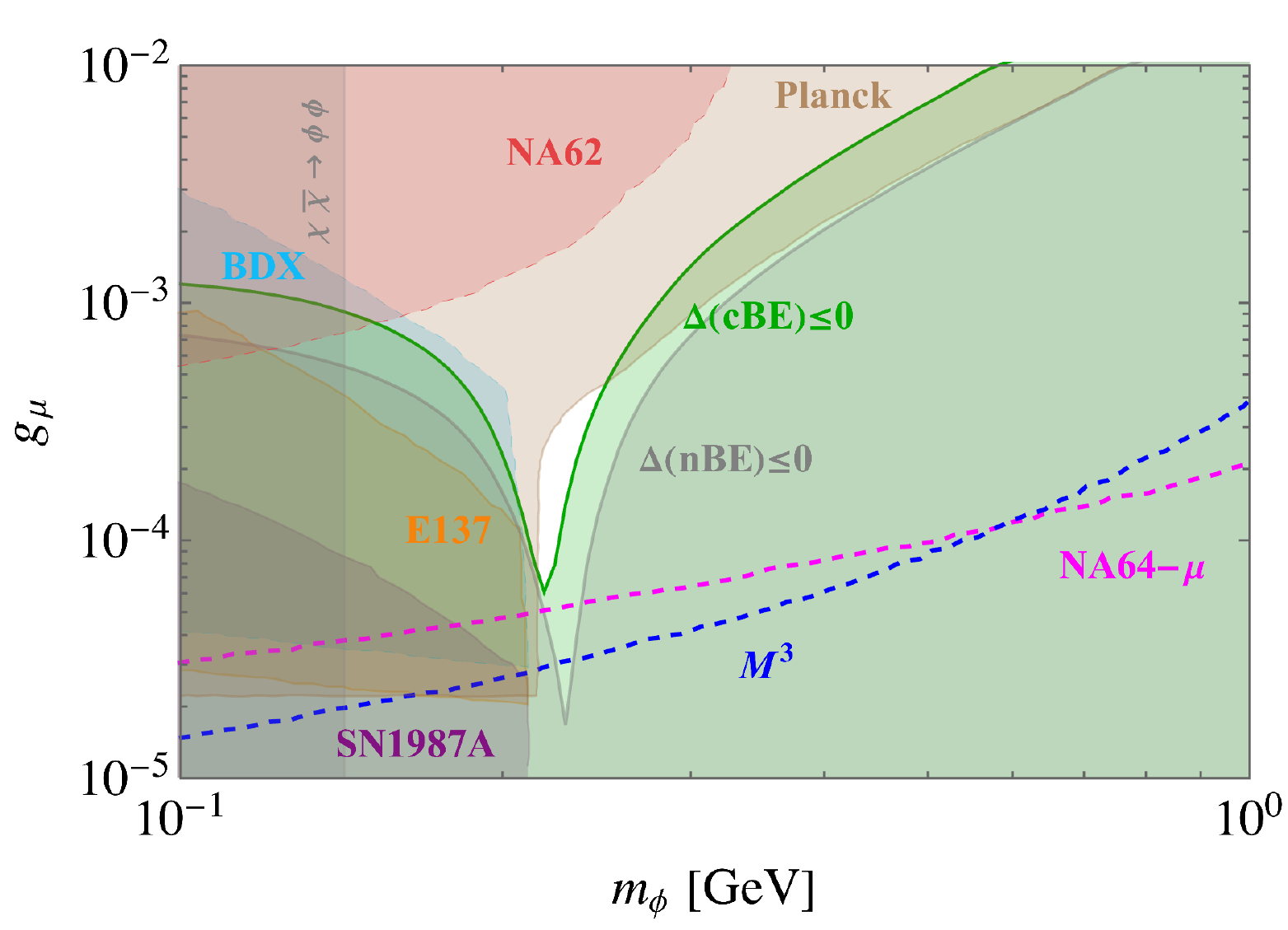}
  \includegraphics[width=0.475\textwidth]{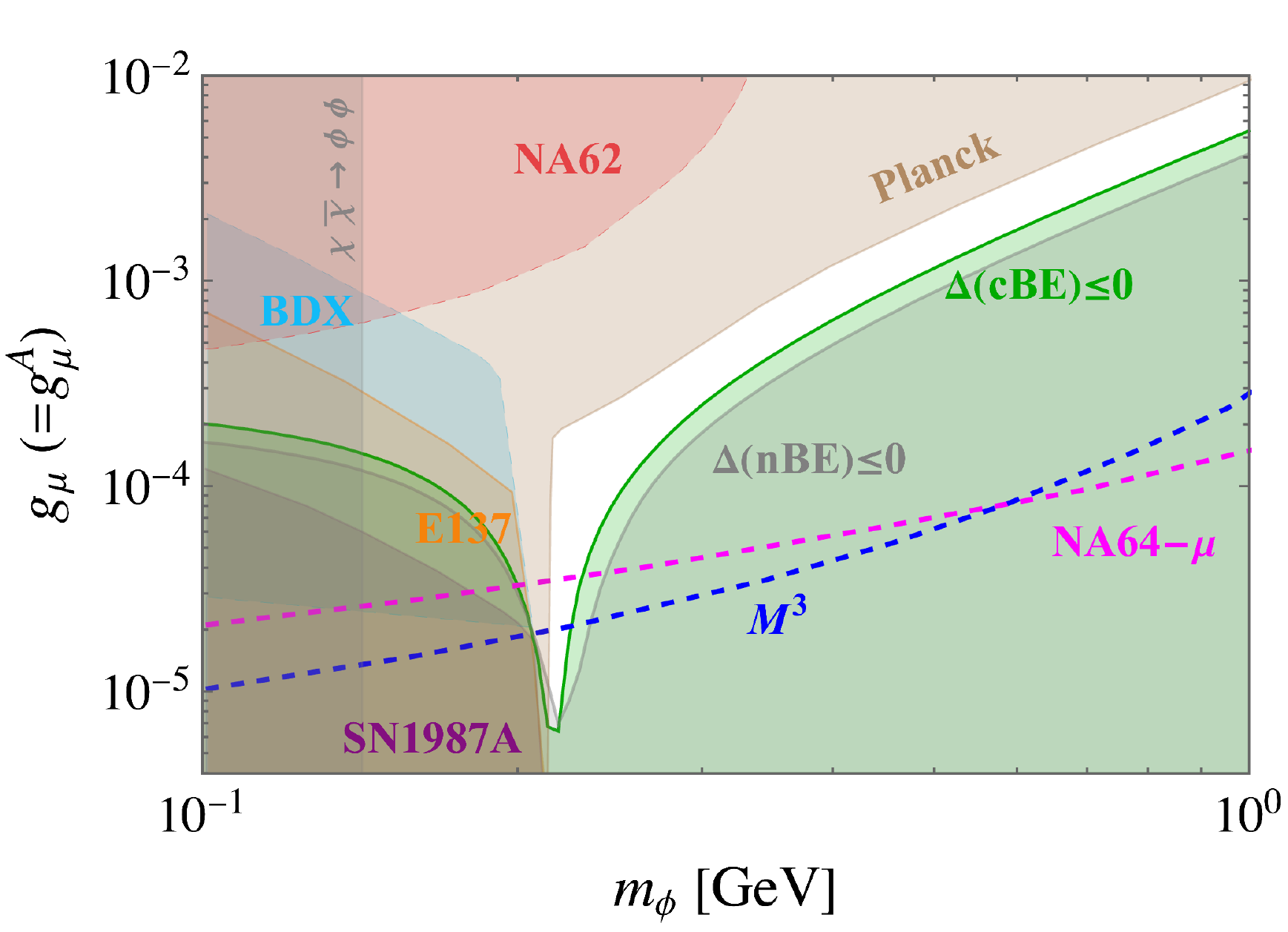}
  \caption{Results and constraints on forbidden channel $\chi\bar\chi\to\mu^+\mu^-$. 
  The green line represents the parameter boundary of the forbidden DM region with $ m_\mu/m_\chi=1$; and the green shaded region stands for the non-forbidden space that $m_\chi>m_\mu$. The rest space is allowed for the forbidden annihilations to get the correct DM relic density. 
  The gray region  labeled $\Delta ({\rm nBE}) \leqslant0$ is obtained in the {\rm nBE} approach. Also, the non-forbidden space that is dominated by the $\chi\bar\chi\to\phi\phi$ process is shown. 
  The brown region displays the Planck bounds on the energy injection process \cite{Slatyer:2015jla}. 
  The orange region is excluded by the E137 electron beam-dump experiment \cite{Bjorken:1988as}. 
  The purple region is excluded by energy loss process of SN1987A \cite{Kamiokande-II:1987idp,Jaeckel:2017tud,Dolan:2017osp}.
 The projected sensitivities for future beam-dump experiments from BDX \cite{Bondi:2017gul} (light blue), M$^3$ \cite{Berlin:2018bsc, Kahn:2018cqs} (dashed blue line), $\mathrm{NA62}$ \cite{NA62:2017rwk} (red), and {NA64}-$\mu$ \cite{Chen:2018vkr,Gninenko:2014pea, Gninenko:2001hx}   (dashed magenta line) are also shown here.}
  \label{Fig:cons-mumu}
\end{figure*}

We first study the $\mu^+\mu^-$ forbidden channel. The numerical results are shown in Fig.~\ref{Fig:cons-mumu}. We set $g_\mu^A=0$ for the left panel, while $g_\mu^A=g_\mu$ for the right panel. The different choices of the couplings to muons do not strongly affect the phenomenology. 

To elaborate on our findings, we start by searching the parameter boundary of the forbidden annihilation, in which $\Delta\to0$ and the right relic density are fulfilled. 
The green line in the plots shows the boundary of the {\bf cBE} approach; the green shaded region, denoted by the expression $\Delta(\rm {\bf cBE}) \leqslant 0$, represents the unforbidden space.
In the remaining parameter space, where the forbidden annihilations dominate the DM depletion, $m_\chi$ is chosen at each point to match the correct relic density. 
The required coupling becomes larger when  $m_\mu/m_\chi>1$, leading to the corresponding curves lying inside the funnel area with a similar shape as the boundary. 
For comparison, we also repeated the parameter boundary of the standard {\bf nBE} approach in the same figure, as shown by the gray lines. The gray regions stand for the non-forbidden regions.

The allowed parameter space of the two approaches is noticeably distinctive. The eKD effect reduces the parameter space, compared to that of {\bf nBE}. 
Or we can say that to satisfy the relic density requirement, the larger coupling $g_\mu$ is required for the {\bf cBE} scenario. 
The reason is straightforward as already pointed out in the last section. 
Around the freeze-out stage, the temperature of DM decreases resulting in reduced kinetic energy and then inefficient forbidden annihilations. To maintain the DM dilution process, a larger coupling is required for $T_\chi < T$.

In the small mediator mass region, the annihilation into pairs of mediators dominates the relic density. We depict it in gray with the label $\chi\bar\chi\to\phi\phi$ as the non-forbidden DM region. For experimental 
constraints, the most crucial parameter is the mediator mass $m_\phi$. In the following, we show the constraints one by one: 

\begin{itemize}
    \item Planck (brown region).
    
    DM annihilations into SM electromagnetically interacting particles could modify the anisotropies of the CMB \cite{Adams:1998nr, Chen:2003gz, Padmanabhan:2005es}. The measurements of the CMB by the Planck satellite\cite{Planck:2018vyg} can thus put robust constraints on such energy injection processes \cite{Slatyer:2015jla}. 
    In this model, the photon pairs can be produced in DM annihilations via a muon loop, leading to the injection into the CMB. We recast the corresponding constraints from Ref. \cite{DAgnolo:2020mpt} which are shown in brown.
    
    As pointed out in Ref. \cite{DAgnolo:2020mpt}, when  $m_\phi$ near $2 m_\chi$ which is also close to but smaller than $2m_\mu$, the annihilation cross section of $\chi\bar\chi\to \gamma\gamma$ is enhanced due to  $\sigma(\chi\bar{\chi}\to\gamma\gamma)\sim1/[(m_\phi^2-4m_\chi^2)^2+m_\phi^2\Gamma_\phi^2]$. So the CMB constraints exclude the most parameter space for $m_\phi<2m_\mu$ and only leave a small part of the allowed room when $m_\phi> 2 m_\mu$.
    
    From Fig.~\ref{Fig:cons-mumu}, we find that the eKD effects narrow down the allowed parameter space that obtains the right amount of DM relic abundance. 
    Especially the CMB constraints exclude the most parameter space with {\bf cBE} in the pure scalar interaction scenario (i.e. $g_\mu^A = 0$).

    \item E137 (for orange region) and BDX (blue region).
    
    Secondary muons are produced from the electron beam-dump experiments, such as SLAC E137 \cite{Bjorken:1988as} and Jefferson Lab BDX experiments\cite{Bondi:2017gul}, which can be used to explore the signals of light scalar emission and the muon-scalar coupling via muon-nucleon scattering process $\mu+N\rightarrow \mu+N+\phi$. The E137 experiment's null result established exclusion limits on the parameter space, and the upcoming BDX experiment can likewise yield a predicted limit.

    In Fig.~\ref{Fig:cons-mumu}, we display, as shaded areas, constraints from the E137 electron beam-dump experiment~\cite{Marsicano:2018vin}, and projections from  BDX~\cite{Bondi:2017gul}. The constraints exclude part of the available parameter space in the ($m_\phi-g_\mu~(g_\mu^A)$) plane.

     \item $\mathrm{NA62}$ (red shaded region).
     
    NA62 is a fixed-target experiment at the CERN Super Proton Synchrotron (SPS) that is dedicated to measurements of kaon rare decays, including projected searches on $K\to \mu\nu\phi$ \cite{NA62:2017rwk}. Such a decay channel is an excellent probe of new light scalars that couples preferentially to muons. 
    Ref. \cite{Krnjaic:2019rsv} has derived the probe sensitivity for this process, which can be used to test our parameter space, as shown in red.

    \item NA64-$\mu$ and M$^3$ (magenta and blue dashed lines).

    Similar to the above beam-dump experiments, NA64-$\mu$~\cite{Chen:2018vkr, Gninenko:2014pea, Gninenko:2001hx} and ${\rm M}^3$~\cite{Berlin:2018bsc, Kahn:2018cqs} are designed to search light scalars in the muon-nucleon scattering process $\mu N\rightarrow\mu N\phi$, using muon beams. It's worth noting that the allowed parameter space can be tested in the coming future.

 \item SN 1987A (purple region).
    
    At last, for the parameter space involving new light scalars, we should also consider the constraints from the observation of supernovae cooling. 
    The most famous constraints arise from the energy loss process via the Primakoff effect $\gamma p\to p \phi$ in SN1987A\cite{Kamiokande-II:1987idp, Jaeckel:2017tud,Dolan:2017osp}. 
    The excluded region is displayed in purple.

\end{itemize}

\subsection{$\chi \bar\chi \to \tau^+\tau^-$}

\begin{figure*}[!htbp]
  \centering
  \includegraphics[width=0.475\textwidth]{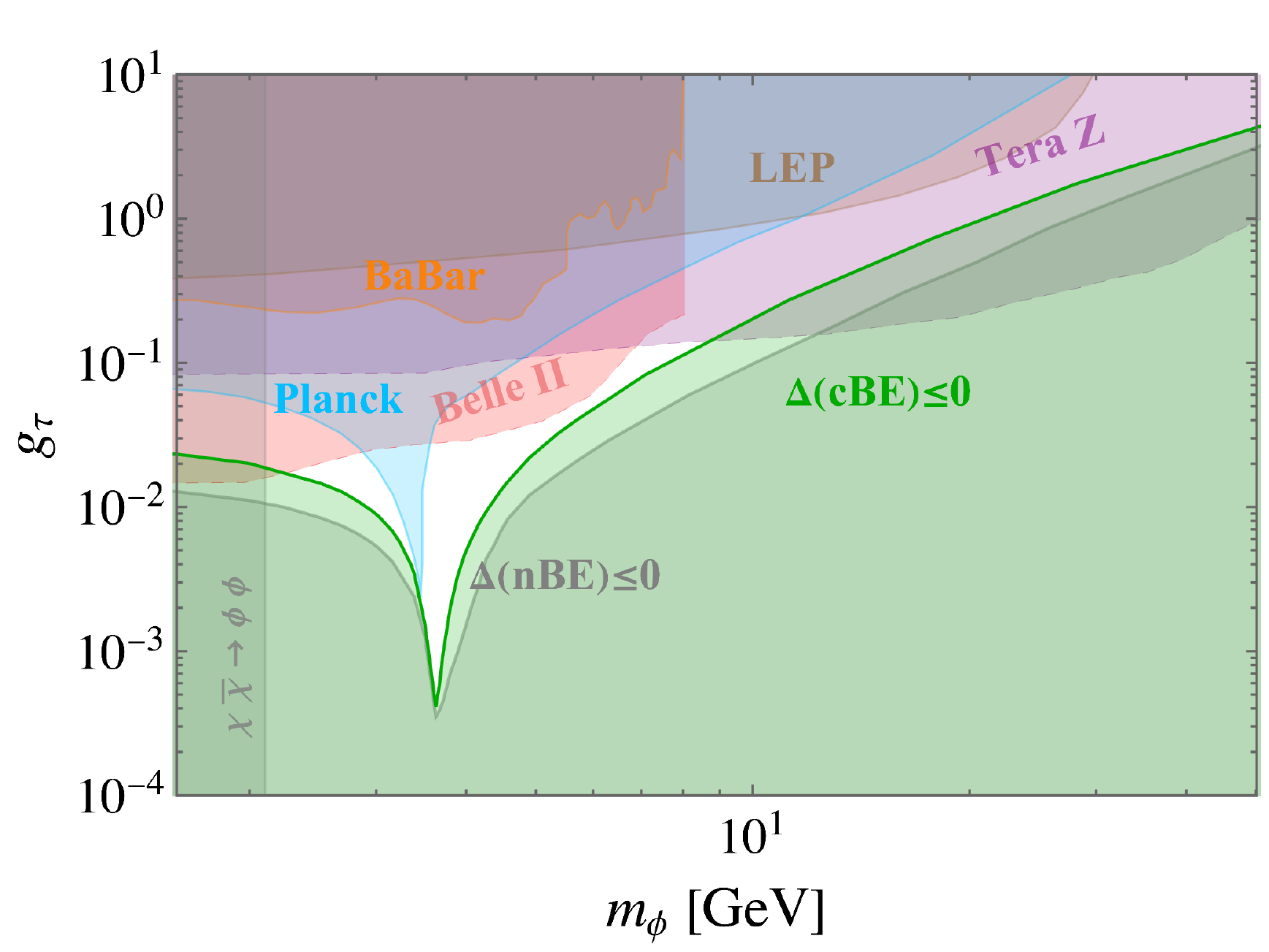}
  \includegraphics[width=0.475\textwidth]{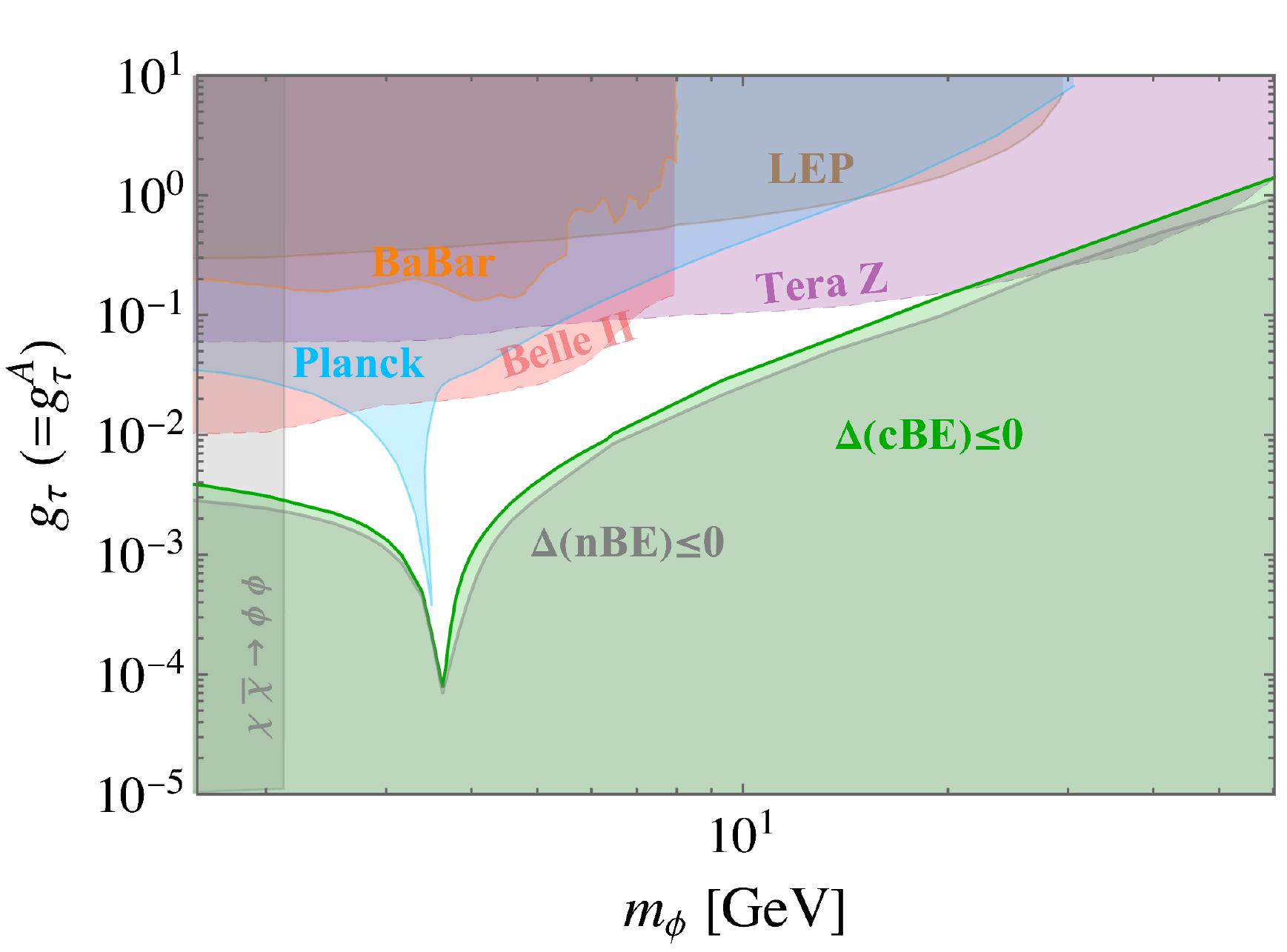}
  \caption{Results and constraints on forbidden channel $\chi\bar\chi\to\tau^+\tau^-$.
  The green and gray shaded regions have the same meaning as in Fig.~\ref{Fig:cons-mumu}, and we use mass ratio $m_\tau/m_\chi=1$.
  We show constraints from 
  searches at BaBar for $e^+e^-\rightarrow\phi\gamma$ \cite{Dolan:2017osp, BaBar:2017tiz, Chen:2018vkr} (orange region), 
  searches at LEP for $Z\rightarrow\bar{\tau}\tau+$MET \cite{Chen:2018vkr, Tanabashi:2018oca} (brown region), 
  and Planck constraints on DM annihilations (blue). 
  We include projections for Belle II \cite{Dolan:2017osp,Belle-II:2010dht} (red region) and future $Z$-factories \cite{Liu:2017zdh, dEnterria:2016fpc, dEnterria:2016sca, CEPC-SPPCStudyGroup:2015csa} (purple region).}
  \label{Fig:cons-tautau}
\end{figure*}

The forbidden annihilation in the di-tau case is discussed here.    
Annihilation to $\tau^+\tau^-$ shares several qualitative features with annihilations to muons. 
The same computations are performed, including the searches for the forbidden annihilation parameter region, the comparison of {\bf cBE} and {\bf nBE} treatments, and the variety of limitations from experimental searches. The results are displayed in Fig.~\ref{Fig:cons-tautau} with different parameter settings. 
Note that the consideration of the experimental constraints is rather different from the $\chi \bar\chi \to \mu^+\mu^-$ case due to the different mediator mass.

\begin{itemize}
    \item BaBar (orange region) and Belle II (red region) 
    
    Searching for $e^+e^-\to \gamma + {\rm invisible}$ at the $e^+e^-$ colliders, such as BaBar and the future Belle II experiments, sets existed and projected constraints on the parameter space. 
    As depicted in Fig.~\ref{Fig:cons-tautau}, the orange line and region show the constraints arising from BaBar~\cite{Dolan:2017osp, BaBar:2017tiz, Chen:2018vkr}, and the red region shows the exquisite sensitivity from Belle II using 50 ab$^{-1}$ integrated luminosity~\cite{Dolan:2017osp,Belle-II:2010dht}, which indicates that a large portion of the viable parameter space will be tested.
    
    \item LEP (brown region) and Tera Z (purple region)
    
    Precision measurements of $Z$ boson decay width can place constraints on the model parameters, namely on the exotic $Z$ decays. In this scenario, $Z\to \tau^+ \tau^- \phi~(\phi\to {\rm invisible})$ contributed to the measured $Z\to \tau^+\tau^-$ width. 
    {Large Electron Positron Collider (LEP)} has set corresponding limits on such channels~\cite{Chen:2018vkr, Tanabashi:2018oca} which are shown in brown.  
    
    Additionally, there have been several proposals for future $Z$-factories to search the same processes \cite{Liu:2017zdh, dEnterria:2016fpc, dEnterria:2016sca, CEPC-SPPCStudyGroup:2015csa}, based on {Circular Electron Positron Collider (CEPC)} and the {Future Circular Collider $e^+ e^-$ (FCC-ee)} for instance. The projections of Tera-Z options (with accumulated $10^{12}$ Z’s ) provide leading sensitivities in the tens GeV range, shown in purple.

    \item Planck (blue region)
    
    The CMB constraints are similar with the $\chi\bar\chi\to\mu^+\mu^-$ forbidden channel. 
    The parameter space at $m_\phi < 2m_\tau$ is much constrained as always, as shown in Fig.~\ref{Fig:cons-tautau}, due to the enhancement of $\chi\bar\chi\to\gamma\gamma$ when $m_\phi \simeq 2m_\chi$.

\end{itemize}


\section{Conclusion}
\label{sec:conclusion}

In the forbidden DM scenario, consideration of the early kinetic decoupling is not only a correction to the DM relic density but an indispensable ingredient. 
In this work, we investigate the early kinetic decoupling effect in forbidden channels, where DM annihilation to SM leptons is kinetically forbidden. Specifically we focus on the $\chi\bar\chi\to\mu^+\mu^-$ and  $\chi\bar\chi\to\tau^+\tau^-$ modes. 
By analyzing the scattering momentum transfer rate, we found the kinetic equilibrium breaks at about $x\simeq20$, which is the same stage of chemical decoupling. So eKD should be taken seriously into account. 
With different benchmark points, we found that there is a cooling phase during the evolution that causes the DM temperature to deviate from the thermal bath and evolve solely. The decreased kinetic energy of DM particles suppresses the forbidden annihilation rate, which gives rise to larger abundances. 
The difference between the {\bf cBE} and traditional {\bf nBE} methods is significant, within part of the parameter space showing a deviation up to an order of magnitude larger, for both $\mu^+\mu^-$ and  $\tau^+\tau^-$ forbidden channels. 
We also considered the experimental constraints from beam-dump experiments, collider searches, and astrophysical observations. 
The viable parameter space in the forbidden DM model has been reduced when using the {\bf cBE} treatment. Most of the parameter space will be tested by the forthcoming experimental searches.


\section*{ACKNOWLEDGEMENTS}
We thank Murat Abdughani for the helpful discussions. 
This work was supported by the National Natural Science Foundation of China under Grants No. 12005180 and No. 12275232 and by the Natural Science Foundation of Shandong Province under Grant No. ZR2020QA083.

\appendix

\begin{widetext}

\section{Cross sections for annihilation and scattering processes}
\label{app:rate}

The annihilation cross section for $\chi\bar\chi\rightarrow l^+l^-$ is  
 \begin{equation}
    \sigma=\frac{\Big(\sqrt{s-4 m_l^2}\big(g^{A2}_{l} s+g_{l}^2 (s-4m_l^2 )\big) \big(g^{A2}_\chi s+g_\chi^2 (s-4m_\chi^2)\big)\Big)}{\Big(16\pi s \sqrt{s-4m_\chi^2}\big((s-m_\phi^2)^2+m_\phi^2\Gamma_\phi^2\big)\Big)}, 
 \end{equation}
where the total decay rate of scalar $\phi$ is calculated as 
\begin{equation}
  \begin{split}
    \Gamma_\phi={} &\frac{1}{16\pi m_\phi}\sqrt{1-\frac{4m_l ^2}{m_\phi ^2}}\times\\ 
              &\big((g_{l}-g^A_l)^2\big(m_\phi ^2-2m_l ^2\big)+(g^A_l+g_l)^2\big(m_\phi^2-2m_l^2\big)- 4m_l^2(g^A_l+g_l)(g_l-g^A_l)\big)\\
              &+\frac{1}{16\pi m_\phi}\sqrt{1-\frac{4m_\chi ^2}{m_\phi^2}} \times\\
              &\big((-g^A_\chi-g_\chi)^2\big(m_\phi^2-2m_\chi^2\big)+(g^A_\chi-g_\chi)^2\big(m_\phi^2-2m_\chi^2\big)+4 m_\chi^2(g^A_\chi-g_\chi)(g^A_\chi+g_\chi)\big)
    \end{split}  
\end{equation}

For the elastic scattering process $\chi\mu^\pm\rightarrow\chi\mu^\pm$, the amplitude is
\begin{equation}
    \vert\mathcal{M}\vert ^2=\frac{(g^{A2}_l t+g_l^2(t-4m_l^2))(g^{A2}_\chi t+g_\chi^2 (t-4m_\chi^2))}{(m_\phi^2-t)^2}. 
\end{equation}

And the momentum transfer rate can be written as, 
\begin{eqnarray}
\gamma= \frac{1}{3 g_{\chi} m_{\chi} T} \! \int \! \frac{\text{d}^3 k}{(2\pi)^3} f_{\cal B}^{\pm}(E_k)\left[1\!\mp\! f_{\cal B}^{\pm}(E_k)\right] \! \! \! \int\limits^0_{-4 k_\mathrm{cm}^2} \! \! \!  \text{d}t (-t) \frac{\text{d}\sigma}{\text{d}t} v\,,
\end{eqnarray}

with

\begin{equation}
  \int_{-4 k_{cm}^2}^0 dt (-t)\frac{d\sigma}{dt}v= \frac{1}{64 \pi k E_k m_\chi^2} \int_{-4 k_{cm}^2}^0 dt (-t) {\vert\mathcal{M}\vert^2}=\frac{1}{64 \pi k E_k m_\chi^2} \times t{\rm Amp}, 
\end{equation}
and 
\begin{equation}
    \begin{aligned}
        t{\rm Amp}&=4 k^2 m_\chi^2\Bigg(\frac{2(g^{A2}_l+g_l^2) (g^{A2}_\chi +g_\chi^2) k^2 m_\chi^2}{(m_\chi^2+m_\mu^2 +2 m_\chi E_k)^2}+{}\\
        &\frac{4(g^{A2}_l +g_l^2 )g_\chi^2 m_\chi^2 +4g_l^2 (g^{A2}_\chi+g_\chi^2) m_\mu^2 -2(g^{A2}_l +g_l^2 )(g^{A2}_\chi +g_\chi^2 ) m_\phi^2}{m_\chi^2 +m_l^2 +2 m_\chi E_k}+{}\\
        &\frac{\big(4g_l^2 m_l^2-(g^{A2}_l +g_l^2)m_\phi^2\big)\big(-4g_\chi^2 m_\chi^2 +(g^{A2}_\chi +g_\chi^2 )m_\phi^2 \big)}{4k^2 m_\chi^2 +m_\phi^2 (m_\chi^2 +m_l^2 +2 m_\chi E_k)}\Bigg)-{}\\
        &\Big(16 g_l^2 g_\chi^2 m_\chi^2 m_\mu^2-8\big((g^{A2}_l +g_l^2) g_\chi^2 m_\chi^2 +g_l^2 (g^{A2}\chi +g_\chi^2 ) m_l^2 \big) m_\phi^2+{}\\
        &3(g^{A2}_l +g_l^2) (g^{A2}_\chi +g_\chi^2 )m_\phi^4\Big)\Bigg(\log m_\phi^2 -\log(m_\phi^2 +\frac{4 k^2 m_\chi^2}{m_\chi^2 +m_l^2 +2 m_\chi E_k})\Bigg). 
    \end{aligned}
\end{equation}

\end{widetext}

\bibliographystyle{apsrev4-1.bst}
\bibliography{lit}

\end{document}